# High Kerr nonlinearity hydrogenated amorphous silicon nanowires with low two photon absorption and high optical stability

C. Grillet,[1,2,*] L. Carletti,[2] C. Monat,[2] P. Grosse,[3] B. Ben Bakir,[3] S. Menezo,[3] J. M. Fedeli,[3] and D. J. Moss[1] **

[1] *Institute of Photonics and Optical Sciences (IPOS) and CUDOS, School of Physics, University of Sydney, New South Wales 2006 Australia*
[2] *Université de Lyon, Institut des Nanotechnologies de Lyon (INL), 69131 Ecully, France*
[3] *CEA-Leti MINATEC Campus, 17 rue des Martyrs 38054 Grenoble Cedex 9, France*
**D.J.Moss current address: School of Electrical and Computer Engineering, RMIT University, Melbourne Vic. Australia 3001
*grillet@physics.usyd.edu.au

**Abstract** We demonstrate optically stable amorphous silicon nanowires with both high nonlinear figure of merit (FOM) of ~5 and high nonlinearity Re($\gamma$) = 1200W$^{-1}$m$^{-1}$. We observe no degradation in these parameters over the entire course of our experiments including systematic study under operation at 2 W coupled peak power (i.e. ~2GW/cm$^2$) over timescales of at least an hour.

### References and links

1. J. Leuthold, C. Koos, and W. Freude, "Nonlinear silicon photonics," Nat. Photonics **4**, 535-544 (2010).
2. M. Foster, A. C. Turner, J. E. Sharping, B. S. Schmidt, M. Lipson, and A. L. Gaeta, "Broad-band optical parametric gain on a silicon photonic chip," Nature **441**, 960-963 (2006).
3. F. Li, M. Pelusi, D. X. Xu, A. Densmore, R. Ma, S. Janz, and D. J. Moss, "Error-free all-optical demultiplexing at 160Gb/s via FWM in a silicon nanowire," Opt. Express **18**, 3905-3910 (2010).
4. H. Ji, M. Galili, H. Hu, M. Pu, L. K. Oxenlowe, K. Yvind, J. M. Hvam, and P. Jeppesen, "1.28-Tb/s demultiplexing of an OTDM DPSK data signal using a silicon waveguide," IEEE Photon. Technol. Lett. **22**, 1762-1764 (2010).
5. B. Corcoran, C. Monat, C. Grillet, D. J. Moss, B. J. Eggleton, T. P. White, L. O'Faolain, and T. F. Krauss, "Green light emission in silicon through slow-light enhanced third-harmonic generation in photonic crystal waveguides," Nat. Photonics **3**, 206-210 (2009). DOI: 10.1038/nphoton.2009.28.
6. B. Corcoran, C. Monat, M. Pelusi, C. Grillet, T. P. White, L. O'Faolain, T. F. Krauss, B. J. Eggleton, and D. J. Moss, "Optical signal processing on a silicon chip at 640Gb/s using slow-light," Opt. Express **18**, 7770-7781 (2010).
7. C. Xiong, C. Monat, A. S. Clark, C. Grillet, G. D. Marshall, M. J. Steel, J. Li, L. O'Faolain, T. F. Krauss, J. G. Rarity, and B. J. Eggleton, "Slow-light enhanced correlated photon pair generation in a silicon photonic crystal waveguide," Opt. Lett. **36**, 3413-3415 (2011).
8. S. Zlatanovic, J. S. Park, S. Moro, J. M. Chavez Boggio, I. B. Divliansky, N. Alic, S. Mookherjea, and S. Radic, "Mid-infrared wavelength conversion in silicon waveguides using ultracompact telecom-band-derived pump source," Nat. Photonics **4**, 561-564 (2010).
9. X. Liu, R. M. Osgood, Y. A. Vlasov, and W. J. Green, "Mid-infrared optical parametric amplifier using Si nanophotonic waveguides," Nat. Photonics **4**, 557-560 (2010).
10. B. Kuyken, X. Liu, G. Roelkens, R. Baets, R. M. Osgood, and W. J. Green, "50 dB parametric on-chip gain in silicon photonic wires," Opt. lett. **36**, 4401- 4403 (2011).
11. B. Kuyken, X. Liu, R. M. Osgood, R. Baets, G. Roelkens, and W. J. Green, "Mid-infrared to telecom-band supercontinuum generation in highly nonlinear silicon-on-insulator wire waveguides," Opt. Express **19**, 20172-20181 (2011).
12. R. K. W. Lau, M. Menard, Y. Okawachi, M. A. Foster, A. C. Turner-Foster, R. Salem, M. Lipson, and A. Gaeta, "Continuous-wave mid-infrared frequency conversion in silicon nanowaveguides," Opt. lett. **36**, 1263-1265 (2011).
13. R. A. Soref, "Mid-infrared photonics in silicon and germanium," Nat. Photonics **4**, 495-497 (2010).
14. B. Jalali, "Silicon photonics: nonlinear optics in the mid-infrared," Nat. Photonics **4**, 506-508 (2010).
15. K. Ikeda, Y. M. Shen, and Y. Fainman, "Enhanced optical nonlinearity in amorphous silicon and its application to waveguide devices," Opt. Express **15**, 17761-17771 (2007).
16. S. K. O'Leary, S. R. Johnson, and P. K. Lim, "The relationship between the distribution of electronic states and the optical absorption spectrum of an amorphous semiconductor: an empirical analysis," J. Appl. Phys. **82**, 3334-3340 (1997).
17. Y. Shoji, T. Ogasawara, T. Kamei, Y. Sakakibara, S. Suda, K. Kintaka, H. Kawashima, M. Okano, T. Hasama, H. Ishikawa, and M. Mori, "Ultrafast nonlinear effects in hydrogenated amorphous silicon wire

waveguide," Opt. Express **18**, 5668-5673 (2010).
18. K. Narayanan, and S. F. Preble, "Optical nonlinearities in hydrogenated amorphous silicon waveguides," Opt. Express **18**, 8998-9905 (2010).
19. S. Suda, K. Tanizawa, Y. Sakakibara, T. Kamei, K. Nakanishi, E. Itoga, T. Ogasawara, R. Takei, H. Kawashima, S. Namiki, M. Mori, T. Hasama, and H. Ishikawa, "Pattern-effect-free all-optical wavelength conversion using a hydrogenated amorphous silicon waveguide with ultra-fast carrier decay," Opt. Lett. **37**, 1382-1384 (2012).
20. K-Y. Wang, and A. C. Foster, "Ultralow power continuous-wave frequency conversion in hydrogenated amorphous silicon waveguides," Opt. Lett. **37**, 1331-1333 (2012).
21. B. Kuyken, S. Clemmen, S. K. Selvaraja, W. Boagaerts, D. Thourhout, P. Emplit, S. Massar, G. Roelkens, and R. Baets, "On-chip parametric amplification with 26.5dB gain at telecommunication wavelengths using CMOS-compatible hydrogenated amorphous silicon waveguides," Opt. Lett. **36**, 552-554 (2011).
22. B. Kuyken, H. Ji, S. Clemmen , S. K. Selvaraja, H. Hu, M. Pu, M. Galili, P. Jeppesen, G. Morthier, S. Massar, L. K. Oxenlowe, G. Roelkens, and R. Baets, "Nonlinear properties of and nonlinear processing in hydrogenated amorphous silicon waveguides," Opt. Express **19**, B146-B153 (2011).
23. H. K. Tsang, R. V. Penty, I. H. White, R. S. Grant, W. Sibbett, J. B. D. Soole, H. P. Leblanc, N. C. Andreadakis, R. Bhat, and M. A. Koza, "Two-photon absorption and self-phase modulation in InGaAsP/InP multi-quantum well waveguides," J. Appl. Phys. **70**, 3992-3994 (1991).
24. O. Boyraz, T. Indukuri, and B. Jalali, "Self-phase-modulation induced spectral broadening in silicon waveguides," Opt. Express **12**, 829-834 (2004).
25. E. Dulkeith, Y. A. Vlasov, X. Chen, N. C. Panoiu, and R. M. Osgood, "Self-phase-modulation in submicron silicon-on-insulator photonic wires," Opt. Express **14**, 5524-5534 (2006).
26. X. Liu, J. B. Driscoll, J. I. Dadap, R. M. Osgood, S. Assefa, Y. A. Vlasov, and W. M. J. Green, "Self-phase modulation and nonlinear loss in silicon nanophotonic wires near the mid-infrared two-photon absorption edge," Opt. Express **19**, 7778-7789 (2011).
27. K. Narayanan, A. W. Elshaari, and S. F. Preble, "Broadband all-optical modulation in hydrogenated-amorphous silicon waveguides," Opt. Express **18**, 9809-9814 (2010).
28. C. Sciancalepore, B. Ben Bakir, X. Letartre, J. Harduin, N. Olivier, C. Seassal, J. M. Fedeli, and P. Viktorovitch, "CMOS-compatible ultra-compact 1.55- μm emitting VCSELs using double photonic crystal mirrors," IEEE Photon. Technol. Lett. **24**, 455 (2012).
29. R. Orobtchouk, S. Jeannot, B. Han, T. Benyattou, J. M. Fedeli, and P. Mur, "Ultra compact optical link made in amorphous silicon waveguide," Proc. SPIE **6183**, *conf. on Integrated Optics, Silicon Photonics, and Photonic Integrated Circuits*, Strasbourg, paper 618304 (2006).
30. K.-Y. Wang, K. G. Petrillo, M. A. Foster, and A. C. Foster, "Ultralow-power 160-Gb/s all-optical demultiplexing in hydrogenated amorphous silicon waveguides," in *Integrated Photonics Research, Silicon and Nanophotonics*, OSA Technical Digest (online), paper IW4C.3 (2012).
31. J. M. Fedeli, M. Migette, L. Di Cioccio, L. El Melhaoui, R. Orobtchouk, C. Seassal, P. Rojo-Romeo, F. Mandorlo, D. Marris-Morini, L. Vivien, "Incorporation of a photonic layer at the metallization levels of a CMOS circuit," in *proceedings of 3rd IEEE International Conf. on Group IV Photonics*, 200-202 (2006).
32. J. M. Fedeli, R. Orobtchouk, C. Seassal, and L. Vivien, "Integration issues of a photonic layer on top of a CMOS circuit," Proc. SPIE **6125**, *conf. on Silicon Photonics*, San Jose, paper 61250H (2006).
33. J. M. Fedeli, L. Di Cioccio, D. Marris-Morini, L. Vivien, R. Orobtchouk, P. Rojo-Romeo, C. Seassal, and F. Mandorlo, "Development of silicon photonics devices using microelectronic tools for the integration on top of a CMOS wafer," Adv. Optical Technol. **2008**, doi:10.1155/2008/412518 (2008).

## 1. Introduction

Single crystal silicon-on insulator (SOI), compatible with computer chip technology (CMOS), has attracted huge interest over the past 10 years as a platform for nonlinear nanophotonic devices for all-optical signal processing [1], primarily because of its ability to achieve extremely high nonlinearities (Re($\gamma$) = $\omega$ $n_2$ / $c$ $A_{eff}$ , where $A_{eff}$ is the waveguide effective area) exceeding 300W$^{-1}$ m$^{-1}$ [2]. All-optical signal processing at bit rates of 160Gb/s [3] to over 1Tb/s [4], third harmonic generation [5] ultra-fast optical monitoring [6] and efficient correlated photon pair generation [7] have now been achieved. However, significant two-photon absorption (TPA) of crystalline silicon (c-Si) at telecom wavelengths is such that its nonlinear figure of merit (FOM = $n_2$ / $\beta_{TPA}\lambda$, where $\beta_{TPA}$ is the TPA coefficient and $n_2$ is the Kerr nonlinearity) is in the range of FOM= 0.3 to 0.5 [1] – much less than ideal for nonlinear optical applications. This has significantly limited the efficiency of these nonlinear devices - the largest parametric gain achieved in the telecom band in c-Si, for example, is only about 2dB [2]. The critical impact of the low FOM of c-Si was dramatically illustrated in recent experiments [8-12] using mid-infrared [13, 14] optical pumps at wavelengths above the TPA threshold wavelength of 2.2μm where c-Si's FOM is much higher (up to ~ 4 [9]). This resulted for instance, in extremely high parametric gain over 50dB [10], super continuum generation [11], and high CW FWM conversion efficiency [12] in Si wires. However, the low FOM of c-Si in the telecom band is a fundamental material property that cannot be improved.

It was recently suggested [15] that amorphous silicon could represent a promising alternative to crystalline silicon due to its expected lower nonlinear absorption resulting from the larger electronic bandgap of amorphous silicon compared to c-Si [16]. Although initial nonlinear measurements in amorphous silicon devices yielded a FOM no better than c-Si (~0.5) [17, 18], recent demonstrations have confirmed the possibility of increasing the FOM from around 1 [19, 20] to as high as 2 at telecommunication wavelengths [21, 22], allowing very high parametric gains of over +26dB over the C-band [21] to be achieved. However, to date a key drawback for this material has been a lack of stability, resulting in a dramatic degradation in performance over relatively short timescales (on the order of a few tens of minutes) [22]. Unless this problem can be solved, despite its very promising nonlinear performance, amorphous silicon could well be in danger of becoming an academic curiosity.

In this paper, we demonstrate amorphous silicon nanowires combining high nonlinear FOM, high nonlinearity and good material stability at telecom wavelengths. We observe no degradation in the nonlinear characteristics at comparable power levels, and over longer timescales than those studied in [22]. We experimentally measure self-phase modulation and nonlinear transmission, fit to standard theory, in order to estimate both a high nonlinear FOM and nonlinear Re($\gamma$) factor.

## 2. Experiment

The hydrogenated amorphous silicon nanowires (a-Si:H) waveguides were fabricated in a 200mm CMOS pilot line at CEA-LETI, France. The a-Si:H film was deposited by plasma enhanced chemical vapor deposition (PECVD) at 350°C on 1.7µm oxide deposited on a bulk wafer. After deposition of a silica hard mask, two steps of 193nm DUV lithography and HBr silicon etching were used to define grating couplers that were well aligned with serpentine waveguides with varying lengths (1.22 cm to 11 cm). The fabricated waveguides are ~ 220nm in thickness and ~ 500nm in width. The cross-section is shown in Fig. 1. A 500nm oxide was deposited to provide an upper cladding. The group velocity dispersion for the TE mode confined within a 500nm×220nm nanowire was calculated with FEMSIM, yielding an anomalous second-order dispersion parameter $\beta_2=-4.2\times10^{-25}$ s$^2$/m at $\lambda=1550$nm.

Figure 1 shows a schematic of the experimental setup used for the measurement of both the linear and nonlinear propagation characteristics of our a-Si:H nanowires. A mode-locked fiber laser with near transform limited ~1.8ps long pulses at a repetition rate of 20MHz at 1550nm was coupled into the TE mode of our a-Si:H nanowires via in-plane gratings. The fiber to waveguide coupling loss per coupler was ~ 10.6dB and 12.4dB per entry and exit, respectively, which was higher than expected due to the grating couplers not being optimized. The propagation loss of the TE mode was measured to be ~ 4.5 dB/cm, via a cut-back method on serpentine waveguides with lengths varying from 1.22 cm to 11 cm. A comparison with linear measurements performed on 1.3mm long straight nanowires yielded a loss contribution due to the bends (10 µm radius) of about 4dB, i.e. on the order of 0.04dB/ bend.

To determine the nonlinear parameters of our waveguides, we performed a series of self-phase modulation (SPM) measurements in a 1.22 cm long nanowire, with a coupled peak power up to ~3W. The output spectrum was then measured as a function of input power.

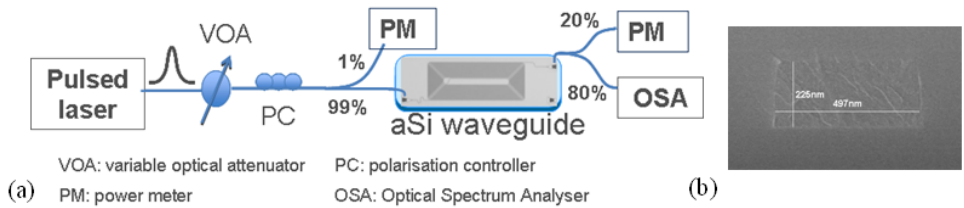

Fig. 1. (a) Experimental setup for observing SPM in a-Si:H waveguide. (b) SEM cross-section of the a-Si:H nanophotonic wire embedded in silica.

## 3. Results and discussion

Figure 2(a) shows the measured output spectra of the pulses for three coupled peak powers, while Fig. 3(a) shows the experimental contour plots of the output spectra as a function of coupled peak power ranging between 0.03W and 3W. Strong spectral broadening is observed, which is the signature of self-phase modulation of the pulse propagating along the nonlinear nanowire. Note that for higher coupled powers (>1.8W), the spectral broadening of the pulse was limited by the spectral transfer function of the grating couplers, which have a 3dB bandwidth of around ~35nm centered near 1550nm. In Fig. 4, we plot the inverse of the waveguide transmission $T$ as a function of coupled peak power. The effect of the couplers can be seen at high powers on this curve as well. As the spectral broadening induced by SPM increases with the power, the lateral sidebands of the output pulse spectrum are attenuated by the transfer function of the couplers, causing a drop in the transmission (i.e. a more rapid increase in $1/T$). Due to this limitation, the $1/T$ curve is fit at low powers up to ~1.2W with the following linear equation valid in the presence of both linear propagation loss and TPA [23]:

$$\frac{1}{T} = \frac{P(0)}{P(L)} = 2\,\mathrm{Im}(\gamma)L_{eff}e^{\alpha L}P(0) + e^{\alpha L} \quad (1)$$

where $P(0)$ and $P(L)$ are the optical peak power at the entrance (coupled peak power) and at the end of the waveguide, respectively, $\alpha=179$/m is the equivalent linear propagation loss including the bend loss contribution, $\mathrm{Im}(\gamma)=\beta_{TPA}/(2A_{eff})$ is the imaginary part of the $\gamma$ nonlinear coefficient due to TPA, $L$ is the physical length and $L_{eff}$ the effective length reduced by the linear propagation loss through $L_{eff}=(1-e^{-\alpha L})/\alpha$. This allows us to extract $\mathrm{Im}(\gamma)=18$/W/m±5%. Assuming a nonlinear modal area $A_{eff}$ of $0.07\,\mu m^2$, we infer a TPA coefficient $\beta_{TPA}$ equal to $0.25\times 10^{-11}$m/W±5%. The uncertainty in this parameter, generally dominated by the uncertainty of the coupled power [18], is reduced here by comparing the nonlinear measurements when coupling to the waveguide from one direction and the other. This yields a difference of ~1.8dB±0.15dB between the left and right hand side insertion loss.

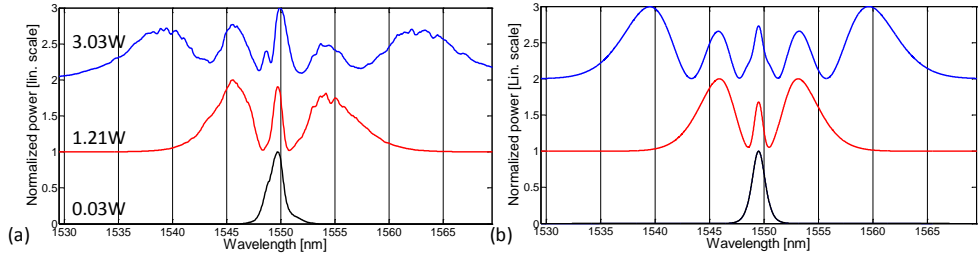

Fig.2. Output spectra for 0.03W, 1.21W and 3.03W coupled peak power - (a): Experiment, (b): Simulation. The curves are normalized and shifted upwards with increasing powers for clarity.

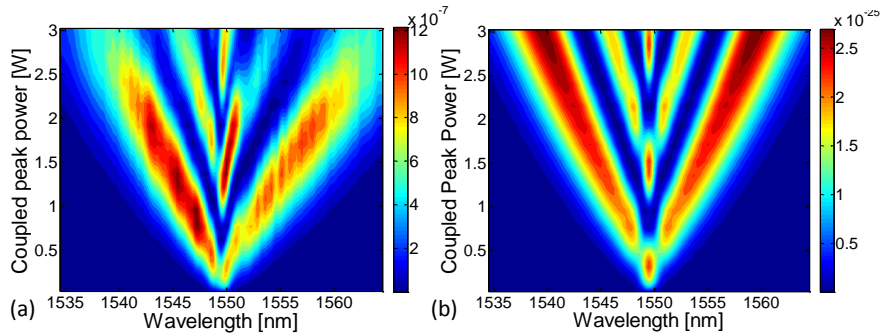

Fig. 3. (a) Experimental and (b) theory 2D plots showing the spectral broadening of the output pulse spectra vs coupled peak power. Note the linear intensity scale at the right is relative.

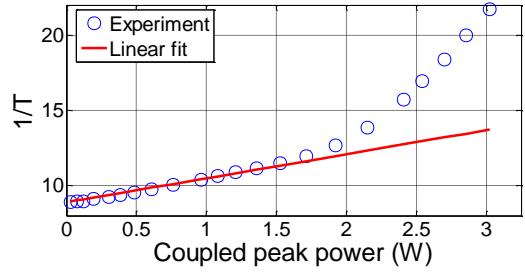

Fig. 4. Inverse of the measured waveguide transmission versus coupled peak power (circles) along with a linear fit at low power.

Split-step-Fourier method simulations were then performed to solve the nonlinear Schrödinger equation governing the propagation of the picosecond optical pulse in the nonlinear waveguide, in the presence of second-order dispersion $\beta_2$ and TPA. The impact of dispersion was negligible, as expected for a $\beta_2 = -4.2 \times 10^{-25} \, \text{s}^2/\text{m}$, associated with a dispersion length exceeding 7m for the 1.8ps pulses, i.e. well over the physical length of the waveguide. Figures 2(b) and 3(b) show the output spectra resulting from these numerical simulations, showing a good agreement with the measurements, when taking the TPA contribution stated above and a nonlinear waveguide parameter $\text{Re}(\gamma) = 1200/\text{W}/\text{m}$, associated with a refractive index $n_2 = 2.1 \times 10^{-17} \, \text{m}^2/\text{W} \pm 5\%$. The good agreement is highlighted in Fig. 5, by directly comparing the nonlinear phase-shift as a function of coupled power for both experiment and simulations. Note that the low TPA of a-Si:H is also reflected in the absence of any blue shift in the output spectra. The latter is a signature of free carriers generated by TPA over the pulse duration and is generally observed in c-Si waveguides for near-infrared picosecond pulses exhibiting similarly high nonlinear phase shifts [24-26]. Although we did not observe any signature at all of free carriers in our material, we note that in principle free carriers can be important and this critically depends on the material system used. For instance all-optical modulation based on free carrier absorption [27] in a-Si:H waveguides has been demonstrated and free carrier lifetimes of ~ps have been measured in [19].

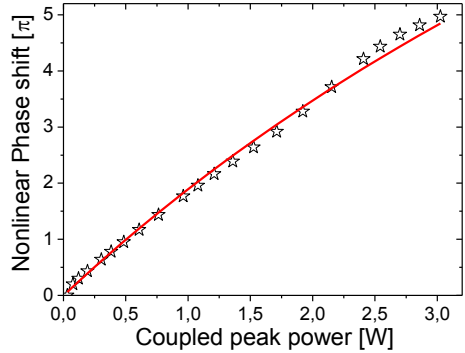

Fig. 5. Nonlinear phase shift versus coupled peak power extracted from experiment (stars) and simulations (red line).

The nonlinear index $n_2$ we obtain for this a-Si:H nanowire is about 4 times higher than c-Si. Together with the inferred TPA coefficient of 0.25 cm/GW, we estimate a FOM= 5±0.3 - an order of magnitude higher than c-Si. Table 1 summarizes the results reported so far in the literature. As mentioned, other groups have measured high values for the nonlinearity [18, 20] or low values of TPA [17, 19] but not both, resulting in an overall modest FOM for a-Si:H in the telecom band. Note that in [20], TPA was taken from [15, 18] and not experimentally measured. Others reported reasonably high FOM [21, 22], but it is clear from table 1, that our results represent a significant improvement for the FOM, more than two times higher than any previous reported results, combined with the high nonlinearity.

Table 1: Summary of the nonlinear characteristics reported for a-Si waveguides fabricated by various groups

|  | Ours | [21,22] | [18] | [17] | [19] | [20] |
|---|---|---|---|---|---|---|
| $n_2$ [$10^{-17}$ m$^2$/W] | 2.1 | 1.3 | 4.2 | 0.05 | 0.3 | 7.43 |
| $\gamma$ [W$^{-1}$m$^{-1}$] | 1200 | 770 | 2000 | 35 | N/A | N/A |
| $\beta_{TPA}$ [cm/GW] | 0.25 | 0.392 | 4.1 | 0.08 | 0.2 | 4 from [15,18] |
| FOM | 5±0.3 | 2.2±0.4 | 0.66±0.3 | 0.4 | 0.97 | 1.1 |

Most importantly, however, as mentioned a-Si:H has exhibited fundamental instability of its nonlinear characteristics when subjected to optical signals in the telecom band [22]. In experiments with similar pulsed laser characteristics to ours, a significant degradation in the modulation instability sidebands was observed after only a few minutes [22]. Even though this could be reversed under annealing, the post-annealed material still exhibited the same instability [22]. Figure 6 shows the spectra out of our 1.22cm long waveguide recorded every two minutes when pumped with 2.25W of coupled peak power, corresponding to ~3GW/cm$^2$, over a timescale of 1hr. The spectra show negligible degradation in the SPM spectral pattern. Figures 7(a) and 7(b) display the associated power and RMS spectral broadening of the output pulse as a function of time. Both show random fluctuations of a few percent over the course of the exposure, which were more likely due to variations in the coupling rather than any clear trend associated with material changes induced by the exposure. Further, we observed no degradation in the parameters over the entire course of our experiments, which at times were conducted at peak powers of up to 3W (~4GW/cm$^2$), primarily limited by the coupler insertion loss. We note that our material has also exhibited stable operation in the context of a hybrid integrated laser [28].

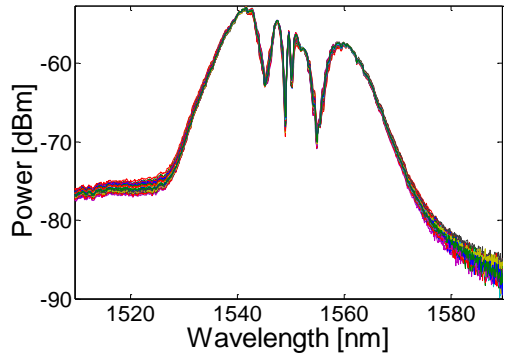

Fig.6. Output spectra as a function time for a coupled peak power of 1.5W. The spectra are recorded every 2min over 1 hour.

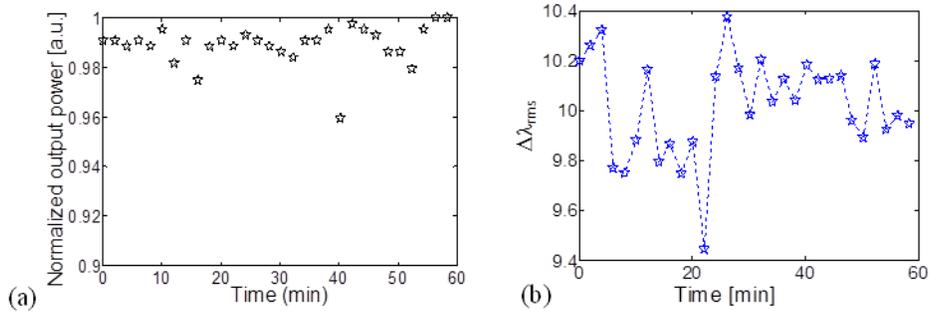

Fig. 7. (a) Normalized power and (b) RMS spectral broadening of the output pulse measured under 2.25W coupled peak power launched over one hour duration.

Finally, a-Si:H is deposited at low temperature and is compatible with standard CMOS technology. a-Si:H based photonics can be considered as an alternative technology to SOI for integrating photonic functions on CMOS [29-33]. Our results raise the prospect for a-Si:H to provide a truly practical and viable, high performance, platform for nonlinear photonic applications in the telecommunications wavelength window.

## 4. Conclusion

We demonstrate a-Si:H nanowires with both high nonlinearity of Re($\gamma$) = 1200 $W^{-1}m^{-1}$ and nonlinear FOM of ~5. We observe no degradation in these parameters under systematic studies at 2.2W coupled peak power over timescales up to an hour, nor did we observe any degradation in the nonlinear parameters over the entire course of our experiments, at times up to 3W coupled peak power. These results represent the first report of simultaneously high FOM, nonlinearity and stability under ~watts coupled peak power operation in a-Si:H and as such potentially pave the way for a-Si:H to offer a viable high performance nonlinear platform for all-optical devices in CMOS compatible integrated nanophotonic circuits, operating in the telecommunications window.

**Acknowledgements**

We acknowledge financial support of the European Union through the Marie Curie program (ALLOPTICS) and the Australian Research Council (ARC).